\documentclass[a4paper, 10pt, conference]{ieeeconf}      % Use this line for a4 paper

\IEEEoverridecommandlockouts                              % This command is only needed if 
\usepackage{graphics} % for pdf, bitmapped graphics files
\usepackage{mathptmx} % assumes new font selection scheme installed
\usepackage{times} % assumes new font selection scheme installed
\usepackage{amsmath} % assumes amsmath package installed
\usepackage{amssymb}  % assumes amsmath package installed

\usepackage{todonotes}
\let\opentask\todo % redefining todo{} as todo[inline]{} to appear in the text not in margins
\renewcommand{\todo}[1]{\opentask[inline,color=red!40]{#1}}

\newtheorem{theorem}{Theorem}%[section]
\newtheorem{corollary}{Corollary}%[section]
\newcommand{\qed}{\hfill $\blacksquare$}

\title{\LARGE \bf
On the Dynamism of User Rejections in Mobility-on-Demand Systems
}

\author{Florian Dandl$^{1}$, Roman Engelhardt$^{1}$, and Klaus Bogenberger$^{1}$% <-this % stops a space
\thanks{$^{1}$Florian Dandl, Roman Engelhardt, and Klaus Bogenberger are with the Chair of Traffic Engineering and Control, Technical University of Munich, 80333 Munich, Germany. \newline
Corresponding author: {\tt\small florian.dandl@tum.de}}%
}

\begin{document}

\maketitle
\thispagestyle{empty}
\pagestyle{empty}

%%%%%%%%%%%%%%%%%%%%%%%%%%%%%%%%%%%%%%%%%%%%%%%%%%%%%%%%%%%%%%%%%%%%%%%%%%%%%%%%
\begin{abstract}

Mobility-on-demand (MoD) systems, especially ride-hailing systems, have seen tremendous growth in recent years. These systems provide user-centric mobility services, whose users expect a high level of convenience. Waiting for a response after an app request and eventually learning after a long period of time that no vehicle is available is hardly acceptable. This study investigates the use-case where users should be served within a certain maximum waiting time. Under certain assumptions, which are reasonable for an attractive MoD business model, it can be shown that an operator using dynamic state optimization can communicate a rejection to users after the first iteration, thereby eliminating unnecessary waiting time before these users would leave the system. Furthermore, early operator rejections reduce the dimension of subsequent customer-vehicle assignment problems, thereby saving computational resources and solving the problems faster. In turn, this allows shorter re-optimization cycles and once again faster responses, i.e. a better user experience.
\end{abstract}

%%%%%%%%%%%%%%%%%%%%%%%%%%%%%%%%%%%%%%%%%%%%%%%%%%%%%%%%%%%%%%%%%%%%%%%%%%%%%%%%
\section{Introduction}
The growth of mobility-on-demand (MoD) services over recent years gave them a non-negligible market share in the urban transportation market. It can be expected that the cost reductions with the introduction of autonomous vehicles will increase the demand for these systems even further. As a consequence, the body of literature about the operation and impact of these services has grown rapidly (see e.g. Narayanan et al.~\cite{Narayanan.2020} for a recent review about autonomous MoD systems). The operation of MoD systems requires solving vehicle routing problems, which have been studied for more than four decades now~\cite{Psaraftis.2016b}.

Psaraftis et al.~\cite{Psaraftis.2016b} use \textit{Ability to Reject Customers} as one of several criteria to categorize the literature. Hyland and Mahmassani identified three categories for the acceptance/rejection decision model: \textit{No Decision}, which means that a request is binding, i.e. customers wait indefinitely and an operator has to serve them eventually, \textit{Fleet Manager Decision}, where the operator can reject a user request, and \textit{Customer Decision}, where a user leaves the system because they did not receive an offer or the offer is not adequate.  Furthermore, the taxonomies further include a criterion for the time constraints, which can be distinguished between no time constraints, soft time constraints (i.e. penalty terms in the objective function for earliness/lateness) and hard time constraints. Time constraints are closely related to the acceptance/rejection model: hard constraints are required for a rejection (either by operator or user), while soft constraints aim at finding user-centric solutions in a system with binding requests.

All three acceptance/rejection models have been applied in recent literature purposefully. On the one hand, there should be a guarantee for both users and operators that requests are binding, if MoD is part of public transportation system (with join design~\cite{Pinto.2019b}) or a higher-level mode-choice model determined that a user should be served by an MoD system~\cite{Horl.2019b,Oke.2020}. It is also easier to compare performance of operating strategies if the share of served users does not have to be part of a multi-criterion evaluation~\cite{Hyland.2018,Dandl.2019} because the selection of requests to serve does impact the other fleet metrics. On the other hand, it cannot be expected that users really wait indefinitely. For the comparison of operational strategies, strict rules for acceptance, i.e. hard time constraints, are more tractable than probabilistic user acceptance models~\cite{Engelhardt.2019,AlKanj.2020} or integrated mode-choice models~\cite{Wilkes.2021,Dandl.2021b}.

The hard constraints can be applied at different times. The information process, i.e. the user-operator interaction model, are decisive for the acceptance/rejection model. Without any information, users will wait until their maximum waiting time constraint is reached and retract their request then. In contrast, there might be boundaries on how much time the operator has to respond before users drop their request~\cite{Dandl.2019c,Yu.2020}. In this response, the operator could reject a user, or the user might retract the request as the offered level-of-service (pick-up and possibly drop-off time) are not within the time-constraints. It should be noted that from a system performance standpoint it does not matter, whether this is modeled as an operator rejecting a user, whose requested time constraints cannot be satisfied, or a user rejecting an operator offer, because the offer does not satisfy these constraints. Therefore, this paper distinguishes only between two cases: a user walking away after their respective maximum waiting time expires and an operator rejecting a request with known time constraints.

The contribution of this study is the derivation of a theoretic connection between the two cases of operators rejecting a request early and users leaving the system after waiting for a certain amount of time. This connection is valid under a certain set of MoD operating policies (hailing and pooling) with hard user time constraints. Section~\ref{sec:mod_definition} defines the problem and identifies the assumptions to the MoD operator control that are required for the theorem, its proof and a few corollaries in Section~\ref{sec:theorem}. In theory, the proof for pooling inherently proofs the hailing case as well; as the hailing case is more intuitive, both cases are treated separately. Lastly, Section~\ref{sec:conclusion} discusses some implications and future research directions.

\section{Problem Definition}
\label{sec:mod_definition}
\subsection{General Definitions}
Let $G=(N,E)$ be a street network on which an MoD service provider offers service for users. A user request $r$ in the set of all requests $R$ is characterized by the tuple $(x_r^p, x_r^d, \tau_r, \tau^w_r)$ for a hailing system and $(x_r^p, x_r^d, \tau_r, \tau^w_r, \tau^d_r)$ for a pooling system, where $x_r^p \in N$, $x_r^d \in N$, $\tau_r$, $\tau^w_r$, and $\tau^d_r$ are the pick-up location, the drop-off location, the time of the request, the maximum waiting time, and the maximum in-vehicle driving time of user request $r$, respectively. Moreover, user requests are categorized according to their state in following subsets of $R$: $R^u$ contains future requests that have not yet been revealed to the MoD operator, $R^{na}$ is the set of not assigned requests, $R^w$ is the set of assigned requests waiting for pick-up, $R^p$ denotes the set of users that have been picked-up and are on the vehicle. User requests either end up in $R^s$ after they were served and dropped of at their destination or in $R^l$ if they left the system. Users leave the system if their request is rejected by the operator or they walk away after waiting for $\tau^w_r$. The different user sets vary over time, but at any given time are a disjoint union of $R$.

The MoD provider operators a fleet of vehicles $V$. The operator's task is the assignment of routes $\xi_v$ to all vehicles $v \in V$, where a route $\xi_v$ is defined by an ordered list of stops with user pick-up and drop-off locations.

This study assumes that users are impatient regarding the information process. They expect the operator process their request and send assignment/rejection information as soon as possible and value it negatively if they are not assigned. A penalty term $P^{-} \gg 1$ will be used to penalize users not being assigned. In the following, it is assumed that users can be re-assigned until their maximum waiting time is reached; however, the theorem proven in section~\ref{sec:theorem} also holds if re-assignments are restricted. Since the worst-case scenario from a user perspective is that an assignment is communicated but retracted at a later time, there is a penalty $P^{+} \gg P^{-}$ for user requests that are moved from $R^w$ to $R^{na}$.

The prioritized objective of the operator is to minimize the penalties. Additionally, the operator tries to minimize another multi-objective function $\tilde{f}$, which can include, e.g., the total driven distance $d_v$ of all vehicles $v \in V$, realized user waiting times ($t^w_r$) and in-vehicle driving times ($t^d_r$):
\begin{align}
\min f &= \sum_{r \in R^{a,na}}P^+ + \sum_{r: r \in R^l}P^- + \tilde{f} \label{global_objective}\\
\tilde{f} &= \tilde{f}\left( \sum_{v \in V} d_v, \sum_{r \in R^s} t^w_r, \sum_{r \in R^s} t^d_r, ... \right)
\end{align}
where $R^{a,na}$ is the set of all requests that received service confirmation, but were not assigned at a later point in time. These requests are in risk of not being picked-up in time despite an earlier confirmation, which causes a bad reputation for the service. The scale of $\tilde{f}$ is chosen such that the change in the route objectives resulting from assigning a request to a vehicle $\lvert \delta\tilde{f} \rvert \ll P^{-}$ in order to guarantee the dominance of the penalty terms.

\subsection{Dynamic Fleet Control}
In a static vehicle routing problem, in which all requests are known ahead of time, the operator can send a vehicle directly to the next customer pick-up after it picked-up or dropped-of a customer. In a dynamic setting, the next customer request of the static solution might still be unknown to the operator, i.e. in $R^u$. Without advanced future information, an operator will not send the vehicle to the this customer's pick-up location in a dynamic setting and the system performance will generally be better in a static setting.

In the dynamic setting, the operator can make new decisions whenever the system state changes. Since time changes continuously, but the operator cannot make instantaneous decisions and perform the respective actions $A$, this is typically modeled in an event-based (with continuous time) or time-step based (with all events collected in the time steps) setting. For the following, a time-step based setting is chosen, where the time is divided in equidistant time steps $t_i$ with $i \in \{0, 1, 2, ..., N^T\}$.

Without any advanced demand information, an MoD operator can characterize its state $S(t_i)$ by the time $t_i$, the network state $G(t_i)$, the set of active requests $R^t(t_i)$ that can be assigned to its vehicles at time $t_i$, and the state of its vehicles, whereas each vehicle $v$ is described by its position in the network $x_v(t_i)$, its current on-board requests $R^v(t_i)$ and an assigned route $\xi_v(t_i)$. For simplicity, it is assumed that the network state remains constant and can be dropped. 

Operator assignments/actions at time $t_i$ are denoted by $A_i = A(S(t_i))$ and change the system from its state $S(t_i)$ to its post-state $S^P(t_i)$. A state transition function $T^S(S^P(t_i))$ then changes the system to the state $S(t_{i+1})$. In the state transition, vehicles move and are boarded according to their currently assigned routes and new requests are revealed to the operator.

Rewards $R(A_i) = R(A(S(t_i)))$ can be mapped to assignment decisions in order to connect this framework with the global objective. The penalty terms of the global objective are easily transferred to the actions: not assigning a previously assigned request to any vehicle entails a penalty of $P^+$ and rejecting a request or not assigning it before $\tau_r + \tau_r^w$ generates a penalty of $P^-$. To reflect the remaining global objectives, an assigned route $x_\nu$ is rewarded with a value of $\tilde{f}^C(\xi_v)$. This function can contain e.g. route distance, time of pick-up and drop-off of customers that are part of the respective route. 

Dynamic fleet control can therefore be formulated as
\begin{align}
\min_{\{A\}} &~ \sum_{i=0}^{N^T} R(A_i) \\
\text{s.t.} &~ S^P(t_i) = A \left( S(t_i) \right) \\
~ &~ S(t_{i+1}) = T^S \left(S^P(t_i) \right)
\end{align}
where $\{A\}$ denotes the set of actions in all time steps.

The Bellman equation 
\begin{equation}
A^*(S(t_i)) = \text{arg}\min_{A_i} \left( R(A_i) + \gamma \min_{A \setminus \{A_i\}} \sum_{j = i+1}^{N^T} R(A_j) \right)
\end{equation}
with $\gamma=1$ enables a recursive dynamic programming approach to find the optimal action $A^*(S(t_i))$ for all time steps $t_i$ assuming all future state transitions and thereby all future requests are known. This becomes more apparent using the formulation with full dependencies, as e.g.
\begin{equation}
A^*_{i+1} = A^*(S(t_{i+1})) = A^*(T^S(A^*(S(t)))
\end{equation}

In case stochastic knowledge of future demand is available, the value of $\sum_{j = i+1}^{N^T} R(A_j)$ can be approximated. The Bellman equation with the approximation represents an approximate dynamic programming approach, in which the discount factor $\gamma$ is typically set to a value between 0 and 1. This discount factor reduces the weight of future rewards since they are not certain. The operator action/decision space is extended by repositioning assignments for this use case, where repositioning denotes the decision to send a vehicle to a location expecting future demand (rather than an explicit user request).

Without any future demand information, $\gamma$ can be set to $0$ and the operator can only optimize the current state:
\begin{equation}
A^+(S(t_i)) = \text{arg}\min_{A(S(t_i))} R(A_i)
\label{eq:DSO}
\end{equation}
This approach is denoted by \textit{dynamic state optimization}. As mentioned before, it cannot be expected that the repeated application of Eqn.~\ref{eq:DSO} results in the global optimum of the static problem. %This is especially crucial in high-demand situations. Consider following example, where the global static solution would reject a customer $r_1$ in order to serve a customer $r_2$ (with $\tau_{r1} < \tau_{r2}$). However, the dynamic state optimization assigns $r_1$ at time $\tau_{r1}$ because it does not know of $r_2$ yet.

\subsection{Model Definitions}
The remainder of this paper assumes an operator using dynamic state optimization to match users and vehicles, where the global objective is defined as in Eqn.~\ref{global_objective}. Moreover, assigned users are accepted immediately, but request-vehicle re-assignments are still possible.

\paragraph{Revealed Requests, Active Requests for Hailing and Pooling}
The new user requests that are revealed to the operator in the current batch leading up to time $t$ change their status from not revealed to not assigned. Hence, they are shifted from $R^u$ to $R^{na}$. Moreover, they are collected in $R^r(t)$. For the hailing case, $R^{na}$ and $R^w$ represent the set of active requests. Once a user $r$ is picked up, the next stop of the vehicle route is automatically $x^d_r$. Hence, the request does not have to be considered for future assignment processes anymore. For the pooling case, $R^p$ has to be considered for the assigned routes as well as the drop-off stop of a user does not have to follow immediately after the pick-up stop. Nevertheless, the on-board requests are part of the vehicle states and do not have to be assigned to the vehicle again.

\paragraph{VR-Graph, Feasible Vehicles, Competing Users}
An VR-graph connects a vehicle $v$ and an active request $r$ if vehicle $v$ can satisfy all time constraints of related with serving $r$. More specifically, $v$ has to be able to reach $x_r^o$ before $\tau_r + \tau^w_r$ regardless of its current state (i.e. it does not have to be idle) in order for $r$ to be picked-up before $t^w_r$ expires. Additionally, detour constraints of all on-board passengers of $v$ and $r$ have to be considered for pooling. Let $\bar{V}(r,t)$ denote the set of feasible vehicles for a given request $r$ at time $t$. Moreover, $\bar{R}(r,t)$ is defined as the set of competing requests, which is defined as the requests that could be picked-up by $v$ instead of $r$. A critical assumption is that the operator has exact knowledge of travel times for all hypothetical routes. Therefore, a backwards-directed Dijkstra with a search radius of $(t - \tau_r) + \tau^w_r$ can generate $\bar{V}(r,t)$. This search radius for a specific request $r$ becomes smaller over time and no vehicle that is outside of the search radius at time $t$ can enter it at a later time. Moreover, vehicles assigned to other users can even move out of the search radius. Hence, it can be concluded that
\begin{equation}
\bar{V}(r,t' \geq t) \subseteq \bar{V}(r,t)
\label{eq:RV_cont}
\end{equation}

\paragraph{Linear Assignment Problem for Ride-Hailing Systems}
Ride-Hailing systems are defined as systems that only allow the users of one request on board at a time. The batch optimization for this case can be represented by an integer linear assignment problem in the hailing case. With the assumption, that trip durations are longer than the maximum waiting time, there will be no routes $\xi_v$ containing two user pick-ups. Hence, at most one active request has to be assigned to each vehicle (Eqn.~\ref{eq:veh_cond}). This does allow a vehicle en-route to drop-off a user to be assigned to pick up another user subsequently. Clearly, a user only has to be served by one vehicle (Eqn.~\ref{eq:rq_cond}). The primary objectives of the control function $f^C$ can be derived from the global objective (Eqn.~\ref{global_objective}) and the the secondary control objective $\tilde{f}^C (z_{rv}; t)$ will represent the secondary global objectives $\tilde{f}$ for an assignment of request $r$ to vehicle $v$ at time $t$.
\begin{align}
\min_{\{z_{rv}\}} ~ f^C =&  P^+ \sum_{r \in R^{a}} \left( 1 - \sum_{v \in V} z_{rv} \right) \nonumber \\
~ &+ P^- \sum_{r \in R^{na}} \left( 1 - \sum_{v \in V} z_{rv} \right) + \tilde{f}^C(\{z_{rv}\}; t)
\end{align}
subject to
\begin{align}
\sum_{r \in R^{na} \cup R^{a}} z_{rv} \leq 1 ~ ~ ~ &\forall v \in V \label{eq:veh_cond}\\
\sum_{v \in V} z_{rv} \leq 1 ~ ~ ~ &\forall r \in R^{na} \cup R^{a} \label{eq:rq_cond} \\
\left( (t + t_{rv}^w) - (\tau_r + \tau_r^w) \right) z_{rv} \leq 0 ~ ~ ~ &\forall r \in R^{na} \cup R^{a} ~ ~ \forall v \in V \label{eq:wt_cond} \\
z_{rv} \in \{0,1\} ~ ~ ~ &\forall r \in R^{na} \cup R^{a} ~ ~ \forall v \in V
\end{align}
where $z_{rv} = 1$ if request $r$ is assigned to vehicle $v$. Constraint~\ref{eq:wt_cond} ensures that requests are only assigned to vehicles if they can be picked-up in time, i.e. the waiting time related to an assignment of request $r$ to vehicle $v$ at time $t$, denoted by $t_{rv}^w$ satisfies $(t + t_{rv}^w) - (\tau_r + \tau_r^w) \leq 0$. The constraint matrix is unimodular, which guarantees that the optimal solution of the relaxed linear problem ($z_{rv} \in [0,1]$) only contains integers. Since constants in the objective function are irrelevant to the optimization problem, the objective function can be reformulated:
\begin{equation}
\min_{\{z_{rv}\}} f^C = - P^+ \sum_{r \in R^{a}} \sum_{v \in V} z_{rv} - P^- \sum_{r \in R^{na}} \sum_{v \in V} z_{rv} + \tilde{f}^C(\{z_{rv}\}; t)
\end{equation}
This can be interpreted as $P^+$ and $P^-$ being assignment rewards. For a briefer notation and conformity with the ride-pooling case, Eqn.~\ref{eq:wt_cond} is reformulated as
\begin{equation}
z_{rv} \leq F_{rv} ~ ~ ~ \forall r \in R^{na} \cup R^{a} ~ ~ \forall v \in V
\end{equation}
where a $F_{rv} = 1$ represents a feasible assignment and $F_{rv} = 0$ denotes that vehicle $v$ cannot pick-up request $r$ in time.

\paragraph{Assignment Problem for Ride-Pooling Systems}
The assignment problem for ride-pooling requires some more definitions. The solution of the vehicle routing problem to find the best feasible route $\xi_v$ according to the objective function considering all stops, which are necessary for the vehicle $v$ to serve all on-board users and a bundle/set of requests $b$, is denoted by $\epsilon_{vb}$. $\epsilon_{vb}$ represents an edge in a graph connecting vehicle-nodes and request bundles, which serves as the assignment problem graph. If there are no feasible routes for a vehicle to serve a certain bundle, no edge exists in the graph. Let $F_{bv}$ be an indicator whether an edge between bundle $b$ and vehicle $v$ exists, i.e. $F_{bv} = 1$ in case a feasible route exists, else $F_{bv} = 0$.

Furthermore, the bundles can be connected to another layer of the graph containing the request in the respective bundles. Hence, a requests $r$ is connected to all request bundles $b \in B(r)$ that contain this request. Using the assignment reward interpretation, all edges from $r \in R^a$ ($r \in R^{na}$) to bundles in $B(r)$ have the cost $-P^+$ ($-P^-$). Finally the edges $\epsilon(v,b)$ have the costs $\tilde{f}^C(\xi_v)$ of the route $\xi_v$ defining this edge, i.e. the solution to the respective vehicle routing problem. There are various ways to build this graph effectively~\cite{AlonsoMora.2017,Engelhardt.2019,Engelhardt.2019a,Liu.2019,Simonetto.2019}; important for the following is the existence of this three-layered vehicle-bundle-request graph (denoted by RTV-graph by Alonso-Mora et al.~\cite{AlonsoMora.2017}). Finally, let $B$ denote the set of all bundles $b$.

The dynamic state optimization solves the assignment problem
\begin{align}
\min_{\{z_{bv}\}} ~ f^c = & -P^+ \sum_{r \in R^a} \sum_{b \in B(r)} \sum_{v \in V} z_{bv} -P^- \sum_{r \in R^{na}} \sum_{b \in B(r)} \sum_{v \in V} z_{bv} \nonumber \\
& + \tilde{f}^c (\{z_{bv}\};t) \label{eq:p_obj}
\end{align}
subject to:
\begin{align}
\sum_{b \in B} z_{bv} \leq 1 ~ ~ ~ &\forall v \in V \label{eq:p_veh_cond}\\
\sum_{b \in B(r)} \sum_{v \in V} z_{bv} \leq 1 ~ ~ ~ &\forall r \in R^{na} \cup R^{a} \label{eq:p_r_cond} \\
z_{bv} \leq F_{bv} ~ ~ ~ &\forall b \in B ~ ~ \forall v \in V \label{eq:p_feas} \\
z_{bv} \in \{0,1\} ~ ~ ~ &\forall b \in B ~ ~ \forall v \in V \label{eq:p_int}
\end{align}
where $z_{bv} = 1$ if the best feasible route $\xi_v$ to serve all user requests in bundle $b$ is assigned to vehicle $v$. As for the ride-hailing case, the objective function (Eqn.~\ref{eq:p_obj}) has three components with very different scales in order to prioritize assigning previously assigned requests over assigning new requests, which in turn is prioritized over routing related costs ($P^+ \gg P^- \gg \tilde{f}^c$). Eqn.~\ref{eq:p_veh_cond} constrains the solution to assign at most one bundle to each vehicle and Eqn.~\ref{eq:p_r_cond} ensures that each customer is only assigned at most once, i.e. at most one bundle containing a request can be assigned to any vehicle. Constraint~\ref{eq:p_feas} allows only assignment of feasible routes and finally Eqn.~\ref{eq:p_int} is the integral condition.

\begin{figure}[t]
	\centering
    \includegraphics[width=.9\columnwidth]{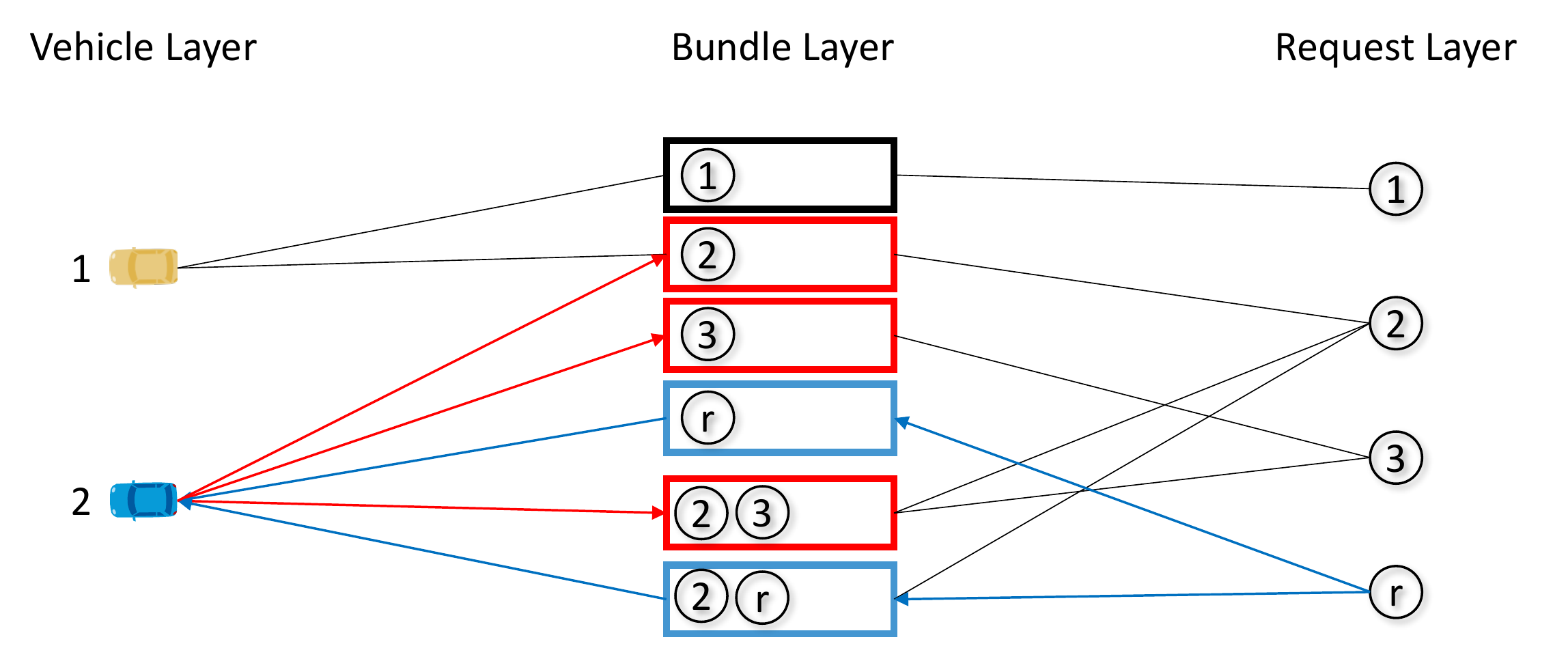}
    \caption{Graphical determination of competing bundles $\bar{B}(r,t)$: following the blue edges returns the set of feasible vehicles from $r$ (vehicle 2). Following the red edges to all bundles not containing $r$ returns the set of competing bundles (highlighted by red boxes).}
    \label{fig:pooling_competing_bundles}
\end{figure}

\paragraph{Competing Bundles} Let $\bar{V}(r,t)$ be the set of feasible vehicles for a given request $r$ at time $t$. Then the set
\begin{equation}
\bar{B}(r,t) := \{b \in B: v \in \bar{V}(r,t) ~ \& ~ F_{bv} = 1 ~ \& ~ r \notin b\}
\end{equation}
is defined as the set of competing bundles. These are all bundles that compete with request $r$ for an assignment to a vehicle. A graphical presentation of this set is given in Fig.~\ref{fig:pooling_competing_bundles}.

\section{Theorem Connecting Rejection by Operator and User Walk-Away}
\label{sec:theorem}
\begin{theorem}
\label{theorem}
Assume an MoD operator as described in section~\ref{sec:mod_definition}. Let $R^r(t)$ be the revealed user requests in the batch leading up to time $t$. For sufficiently large penalty values $P^+ \gg P^- \gg 1$, no request ending in $R^{na}$ after the optimization process will be assigned by any dynamic state optimization before $t + t^{w}$.
\end{theorem}

\begin{corollary}
The sets of requests leaving the system are the same whether the users wait for the maximal waiting time to expire or the operator rejects them after the first batch optimization. Since it is much more convenient for users to know a rejection after a short period, this theorem is useful for practical applications.
\end{corollary}

\subsection{Proof for Ride-Hailing System}

The large penalty terms $P^+$ and $P^-$ cause the solution of the linear assignment problem to be a solution of a maximum priority matching problem
\begin{align}
P1: ~ ~ & \max_{z_{rv}} ~ \sum_{r \in R^a} z_{rv} \label{eq:h_prio1} \\
P2: ~ ~ & \max_{z_{rv}} ~ \sum_{r \in R^{na}} z_{rv} \label{eq:h_prio2}
\end{align}
subject to
\begin{align}
\sum_{r \in R^{na} \cup R^{a}} z_{rv} \leq 1 ~ ~ ~ &\forall v \in V \\
\sum_{v \in V} z_{rv} \leq 1 ~ ~ ~ &\forall r \in R^{na} \cup R^{a} \\
z_{rv} \leq F_{rv} ~ ~ ~ &\forall r \in R^{na} \cup R^{a} ~ ~ \forall v \in V \\
z_{rv} \in \{0,1\} ~ ~ ~ &\forall r \in R^{na} \cup R^{a} ~ ~ \forall v \in V
\end{align}
where the assignment of previously assigned requests $r_a \in R^a$ (Eqn.~\ref{eq:h_prio1}) is prioritized over the assignment of new requests $r_r \in R^r(t)$ (Eqn.~\ref{eq:h_prio2}). In case there are multiple solutions to this maximum priority problem, $\tilde{f}^C(z_{rv}; t)$ determines the request-vehicle pairs. Three cases have to be distinguished:

\paragraph{Trivial} If all requests can be assigned, the theorem is trivial as $R^{na}$ is empty.

\paragraph{No Availability} Let $r \in R^r(t)$ be a request that was not assigned at time $t$ because no vehicle is in the vicinity to pick-up $r$ before $t + \tau^w_r$. Hence, $\bar{V}(r, t) = \varnothing$. Due to Eqn.~\ref{eq:RV_cont} follows that no vehicle will be in the vicinity $\forall t' > t$ and $\bar{V}(r, t + \tau^w_r) = \varnothing$. Hence, the user will leave the system at this time.

\begin{figure}[t]
	\centering
    \includegraphics[width=.95\columnwidth]{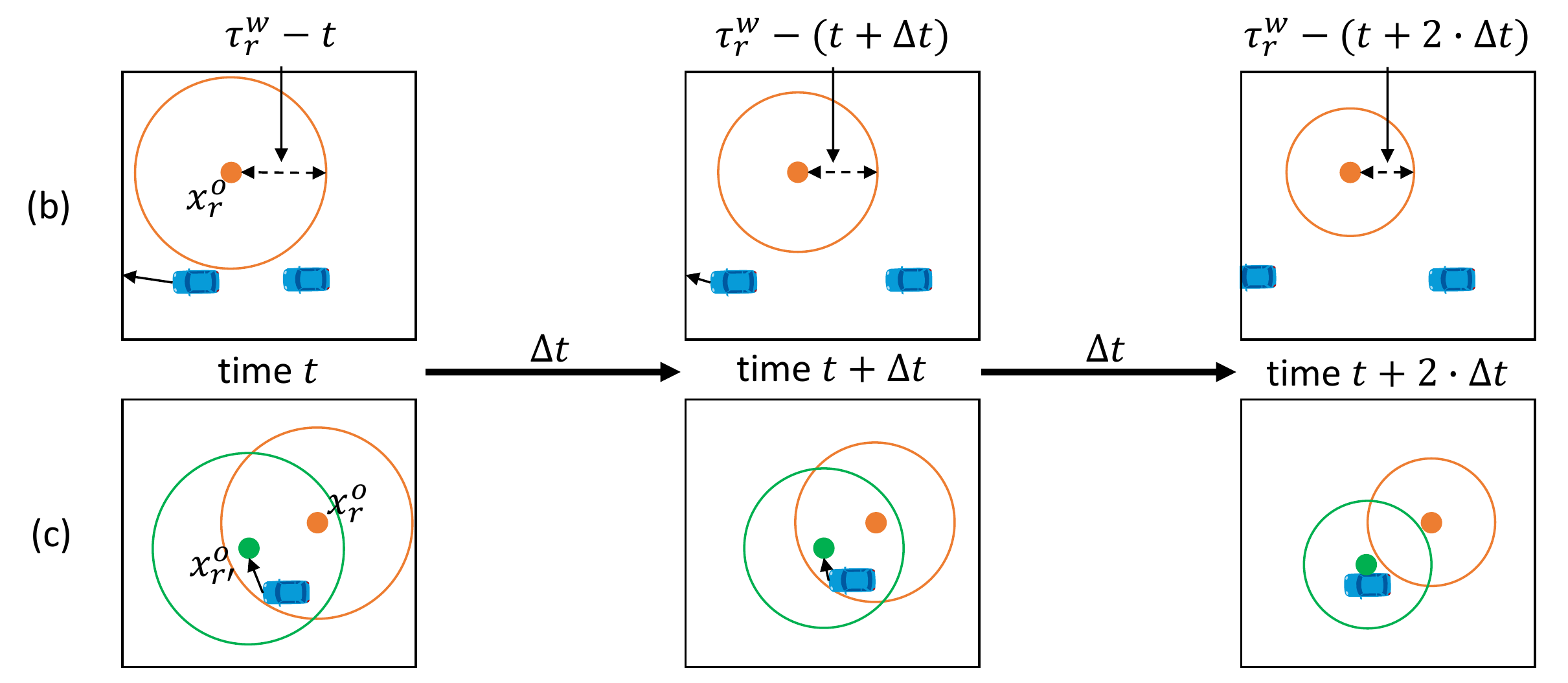}
    \caption{Visualization for cases (b) \textit{No Availability} and (c) \textit{Competing Requests} for ride-hailing assuming Euclidean metric and constant vehicle speed. Vehicles that could serve a request $r$ before a latest pick-up time have to be within a circle and the radius of this circle decreases with the vehicle speed over time. Therefore, no vehicle that is outside the circle can enter at a later time. Moreover, the first step in (c) illustrates that the routing costs $\tilde{f}^c$ related to a possible assignment of a vehicle with $r$ cannot decrease less than those related to this vehicle's current assignment $r'$. It is even possible that the assignment of this vehicle with $r$ becomes unfeasible as shown in the second step of (c).}
    \label{fig:hailing}
\end{figure}

\paragraph{Competing Requests} Let $r \in R^r(t)$ be a request that was not assigned at time $t$ even though $\bar{V}(r, t) \neq \varnothing$. In this case, every vehicle $v \in \bar{V}(r, t)$ is assigned to competing requests in $\bar{R}(r, t)$. The selection of the other requests traces back to (i) them being better positioned with respect to $\tilde{f}^C$ or (ii) them having been previously assigned and the optimization avoiding the penalty $P^+$. Let the next dynamic state optimization take place at $t' = t + \Delta t$. For $\Delta t$, all vehicles $v \in \bar{V}(r, t)$ moved towards their next designated stop determined by the assignments at time $t$. In the worst case the vehicle moves in the opposite direction and $v \notin \bar{V}(r, t')$; in the best case for request $r$ the vehicle moved towards $x_r^o$. However, even in this case the vehicle will not be re-assigned to $r$. Let $\rho_v$ be the request assigned to $v \in \bar{V}(r, t)$ at time $t$. Then the vehicle moved according to the plan and the time to pick-up for $r$ cannot decrease more than the time to pick-up for $\rho_v$. Therefore, following equation holds:
\begin{equation}
\tilde{f}^C(z_{\rho v};t') - \tilde{f}^C(z_{\rho v};t) \leq \tilde{f}^C(z_{rv};t') - \tilde{f}^C(z_{rv};t)
\label{eq:h_obj_dev}
\end{equation}
Hence, the operator has no incentive to swap request-vehicle assignments. Additionally, there exists no matching at time $t'$ that contains all requests in $R^a$ and $r$ as this would have already been the optimal assignment at time $t$. In order to avoid the penalty $P^+$, $r$ will not be part of an optimal assignment at time $t'$ regardless of the existence or attributes of new revealed requests $R^r(t')$. This logic holds until $t + \tau^w_r$ when request $r$ leaves the system. \qed

%%%%%%%%%%%%%%%%%%%%%%%%%%%%%%%%%%%%%%%%%%%%%%%%%%%%%%%%%%%%%%%%%%%%%%%%%%%%%%%%

\addtolength{\textheight}{-3.0cm}   % This command serves to balance the column lengths

%%%%%%%%%%%%%%%%%%%%%%%%%%%%%%%%%%%%%%%%%%%%%%%%%%%%%%%%%%%%%%%%%%%%%%%%%%%%%%%%

\subsection{Proof for Ride-Pooling Systems}
As for the ride-hailing case, the penalty terms require the solution to be an optimum of the respective priority matching problem
\begin{align}
P1: ~ ~ & \max_{z_{bv}} ~ \sum_{r \in R^a} \sum_{b \in B(r)} z_{bv} \label{eq:p_prio1} \\
P2: ~ ~ & \max_{z_{bv}} ~ \sum_{r \in R^{na}} \sum_{b \in B(r)} z_{bv} \label{eq:p_prio2}
\end{align}
subject to
\begin{align}
\sum_{b \in B} z_{bv} \leq 1 ~ ~ ~ &\forall v \in V \label{eq:p_p_obj} \\
\sum_{b \in B(r)} \sum_{v \in V} z_{bv} \leq 1 ~ ~ ~ &\forall r \in R^{na} \cup R^{a} \label{eq:p_s_obj} \\
z_{bv} \leq F_{bv} ~ ~ ~ &\forall b \in B ~ ~ \forall v \in V \\
z_{bv} \in \{0,1\} ~ ~ ~ &\forall b \in B ~ ~ \forall v \in V
\end{align}
where $P1$ is prioritized over $P2$. In case there are multiple solutions to this problem, $\tilde{f}^c$ determines the chosen solution. Three cases have to be distinguished again. The \textit{trivial} and \textit{No Availability} cases are analog to the ride-hailing case.

\textit{Competing Requests/Bundles:} Let $r \in R^r(t)$ be a request that was not assigned at time $t$ even though $\bar{V}(r, t) \neq \varnothing$. In this case, every vehicle $v \in \bar{V}(r, t)$ is assigned to competing bundles in $\bar{B}(r, t)$. All of these bundles have in common that an insertion of $r$ does not generate any feasible routes since the assignment of such route would increase the number of assigned requests and thereby the secondary objective (Eqn.~\ref{eq:p_s_obj}) without changing the primary objective (Eqn.~\ref{eq:p_p_obj}). Let $B^A(t)$ be the optimal solution at time $t$ and $B^b(r, t)$ be the best possible assignments containing request $r$ according to the control objective (Eqn.~\ref{eq:p_obj}). Then $B^A(t)$ either (i) contains more previously assigned requests than $B^b(r, t)$, or (ii) it contains more not assigned requests, or (iii) $\tilde{f}^C (B^A(t)) \leq \tilde{f}^C(B^b(r, t))$. Let $t' = t + \Delta t$, then vehicles moved and users boarded according to the assigned plans. Hence, $B^A(t)$ is still a feasible assignment at time $t'$. It is not necessarily the best assignment at time $t'$ as a matching of a new request and all previously assigned requests (part of $B^A$) might be possible. Nevertheless, the previous assignment is a lower bound for any possible assignment $B^b(r, t')$ including $r$, also including those with new revealed requests. The property $P^+ \gg P^-$ is crucial for this statement. For every possible assignment $B^b(r, t')$ at time $t'$, there exists at least one active request $r'$ that was assigned at time $t$ with $r': r' \in B^A(t) \& r' \notin B^b(r, t')$. However, the assignment of request $r'$ was communicated to the respective user at time $t$, which means that not assigning this user at $t'$ would entail the penalty term $P^+$. Hence, the solution at time $t'$ does not contain $r$. This logic is valid for all time steps $t'' > t$ until the request $r$ times is dropped at time $\tau_r + \tau_r^w$. \qed

\begin{corollary}
All steps of the proof are valid for hailing and pooling systems not allowing request-vehicle re-assignments after the initial assignment is communicated with a user. Hence, the theorem also applies for these systems.
\end{corollary}

\begin{corollary}
\label{heuristic_cor}
The proof is valid for hailing/pooling systems that use heuristics to build the VR/vehicle-bundle-request graph as long as 
\begin{equation}
\forall B^b(r, t'): \exists r': r' \in B^A(t) \& r' \notin B^b(r, t')
\end{equation}
is valid.
\end{corollary}
This condition is satisfied by heuristics that (i) update the costs (such as~\cite{Engelhardt.2019a}) rather than build the graph from scratch (as suggested in\cite{AlonsoMora.2017}) from one time step to the next and (ii) do not allow re-assignments~\cite{Simonetto.2019,Kucharski.2020}.

\section{Conclusion}
\label{sec:conclusion}
This paper establishes a theoretic basis for operators communicating a rejection rather than letting users wait and leave the system when the maximum waiting time expires. This is significant because not knowing whether an MoD operator will provide service for long time is very frustrating from a user perspective. Moreover, the state optimization processes will be computationally more efficient as they do not have to consider all possible routes of rejected requests that anyway will not be assigned in future time steps.

The main requirements/assumptions for the validity of the proof are
\begin{enumerate}
 \item a service guarantee (within the time constraints) is communicated by the operator as soon as a request-vehicle assignment is made; keeping this guarantee is the top priority of the fleet operator
 \item the second priority is to serve as many users as possible
 \item the operator employs dynamic state optimization, e.g. because there is no knowledge about future demand
\end{enumerate}

Future research can study the impact of time-varying and stochastic network travel times, a later commitment to the service guarantee, for which longer response times are traded off for better fleet performance~\cite{Pavone.2020}, and the inclusion of repositioning based on future demand. Likely, simulations will be a necessary tool to analyze the differences between operator rejection and users leaving the system for these studies. Another future research direction can be the design of a system that lets the user set looser time constraints, possibly for a reduced fare.

%%%%%%%%%%%%%%%%%%%%%%%%%%%%%%%%%%%%%%%%%%%%%%%%%%%%%%%%%%%%%%%%%%%%%%%%%%%%%%%%
%\section*{APPENDIX}

%\section*{Acknowledgment}

%%%%%%%%%%%%%%%%%%%%%%%%%%%%%%%%%%%%%%%%%%%%%%%%%%%%%%%%%%%%%%%%%%%%%%%%%%%%%%%%

\bibliographystyle{IEEEtran}
\bibliography{lib}

\end{document}